# Fracture network characterization with deep generative model based stochastic inversion

Guodong Chen[1]; Xin Luo[1]; Jiu Jimmy Jiao[1,*]; Chuanyin Jiang[2]

[1] Department of Earth Sciences, The University of Hong Kong, Hong Kong, China

[2] HSM, University of Montpellier, CNRS, IRD, Montpellier, France

**Corresponding Author(s):** Jiu Jimmy Jiao (jjiao@hku.hk).

## Abstract

The characterization of fracture networks is challenging for enhanced geothermal systems, yet is crucial for the understanding of the thermal distributions, and the behaviors of flow field and solute transport. A novel inverse modeling framework is proposed for the estimation of the fracture networks. The hierarchical parameterization method is adopted in this work. For a small number of large fractures, each fracture is characterized by fracture length, azimuth and coordination of the fracture center. For dense small fractures, fracture density and fractal dimension are utilized to characterize the fracture networks. Moreover, we adopt variational auto-encoder and generative adversarial network (VAE-GAN) and fuse the GAN objective with prior constraint information to capture the distribution of the parameters of complex fracture networks and to satisfy the prior knowledge of fracture fields, thereby mapping the high-dimensional complex parameter distribution into low-dimensional continuous parameter field. Afterwards, relying on the Bayesian framework, ensemble smoother is adopted based on the collected data from hydraulic tomography to reduce the uncertainty of the fracture distribution. Two numerical cases with different complexity are used to test the performance of the proposed framework. The results show that the proposed algorithm can estimate effectively the distribution of the fracture fields.

## 1. Introduction

The in-depth understanding of fractured networks in the bedrocks and the mediated flow and solute transport plays a crucial role in many applied geoscience and energy problems, e.g., heat extraction of geothermal systems, petroleum exploitation, waste disposal [1, 2]. The distribution of fracture network is significant to characterize the behaviors of flow field and solute transport in enhanced geothermal system, and hydraulic fracturing. Commonly used geological monitoring techniques to detect the fracture information are terrestrial laser scanner for outcrops [3], core analysis [4] and micro-seismic interpretation [5, 6]. However, these direct measurements of some key parameters of the fracture networks are difficult as these methods can only obtain limited information of the distribution of fractures. Thus, how to estimate the parameters of the fractures and characterize their distributions have gained increasing attention in recent years [7-10]. Stochastic inversion relying on sufficient dynamic observing data from hydraulic tomography has been proved to be effective and promising to estimate the geometric or statistic parameters of fracture network [11].

As the prerequisite for simulating heat extraction in fractured aquifers, forward simulation is of great significance to inverse modeling [12, 13]. There are three classical fracture modelling methods: equivalent porous medium model [14], dual-porosity model [15], and discrete fracture network model (DFN) [2]. The equivalent porous medium model can simplify the modelling process by equivalent hydraulic approximation [16]. However, the method is oversimplified to portray fractures, causing poor accuracy on flow direction and velocity of groundwater, and impossible to simulate flow behaviors at microscopic scales or in a specific fracture. The dual-porosity model takes into account the combined effects of fractures and matrix using transfer function restricted by pressure, permeability coefficient and geometry [15]. However, the exchange coefficients of different media cannot be obtained through experiments, and it is also challenging to accurately simulate water parcels across the fracture-matrix interface. DFN utilizes unconstructed grid blocks to explicitly express the locations and geometries of the fractures, and models the fracture to portray complex flow behaviors [17]. Solving simulation-based inverse problems generally involves a large amount of forward simulation evaluations. Therefore, how to efficiently and effectively estimate the geometry and other statistic parameters of the fracture network and quantify the uncertainty is indispensable for scientific exploration, engineering work and further decision making.

In the past decades, many algorithms have been employed to inversely estimate the model parameters, such as evolutionary algorithms (EAs) [2, 18, 19], Markov chain Monte Carlo (MCMC) [20] and ensemble Kalman filter (EnKF) [21] and their variants [22]. Under Bayesian framework, the uncertainties of model parameters are quantified by the posterior distribution calculated from prior distribution and likelihood function [23]. EAs, such as genetic algorithm (GA) and differential evolution (DE) [24], are population-based global searching optimization algorithms and have been deployed to solve parameter estimation problems [25]. Nevertheless, inverse problems are normally highly ill-posed and may have many solutions. Thus, EAs are not an effective way to deal with such multi-modal problems. MCMC, first proposed by Metropolis [26], is prevailingly adopted to sample numerous stochastic realizations based on the posterior distributions of parameter fields to quantify the uncertainty of subsurface models. However, it requires numerous time-consuming forward simulations, making it computationally prohibitive to implement millions of model evaluations during the evolutionary search process, particularly for large scale and nonlinear hydrogeological systems. EnKF, firstly proposed by Evensen [21], has been widely applied to atmospheric and geological fields with complex and large-scale inverse problems. Ensemble smoother (ES) [27] improved the updating scheme from sequential to simultaneous. However, ES is incapable to update ensemble of parameters with non-Gaussian or strongly nonlinear properties [28].

Machine learning, especially deep learning, has been widely utilized in the stochastic inversion in the geological problems [8, 29-31]. Commonly used machine learning methods are Polynomial chaos expansion (PCE) [32], Gaussian process (GP) [33, 34], support vector machine (SVM) [35], radial basis function (RBF) [36] and artificial neural network (ANN) [37], etc. PCE was adopted to build surrogate model to replace the real simulation evaluations [38]. Zhang and Lu [39] adopted Karhunen–Loeve expansion to transform the high-dimensional parameter spaces into low-dimensional latent representation. Ruppert, Shoemaker [40] combined EA and RBF to maximize the posterior distribution of parameters efficiently. Wang and Li [41] used GP to replace the real model evaluation and then used MCMC sampling to refine the prediction of posterior density. Recently, deep learning has gained increasing popularity due to its robustness and generalization property on dimensionality reduction and surrogate modeling for uncertainty quantification of model parameters [29, 42]. Deep neural networks (DNN) can predict the underlying relationship between the input parameters and the observing data with complex hierarchy of hidden layers, and also exploit the lower-dimensional feature from the high-dimensional spaces by inherent complex mappings [43, 44]. Laloy, Hérault [45] adopted variational auto-encoder as dimensionality reduction method to alleviate the curse of dimensionality. Mo, Zabaras [46] developed deep convolutional encoder-decoder networks to predict the time-varying saturation and pressure distribution of different permeability field, and further proposed convolutional adversarial auto-encoder as surrogate model to approximate the model simulation of solute transport [47]. Zhang, Zheng [28] adopted deep learning to assist ES to update ensemble for better estimation on non-Gaussian problems. Zhong, Sun [48] used conditional deep convolutional generative adversarial network as the surrogate model to predict the migration of carbon dioxide plume. Zhong, Sun [5] also used cycle generative adversarial neural network on time-lapse seismic reservoir monitoring to improve the reliability of 4-D seismic inversion. Xiao, Zhang [29] proposed a model-reduced adjoint-based optimization workflow integrating convolutional auto-encoder and linear-transition unit on inverse modeling of $CO_2$ sequestration storage. Tartakovsky, Marrero [49] developed a physics-informed deep neural network for learning parameters and constitutive relationships in subsurface flow problems.

The aforementioned studies mainly focus on the stochastic inversions of porous media problems. In recent years, many efforts have been also made on fracture inversion.

Afshari Moein, Somogyvári [50] adopted MCMC sequence to update the DFN iteratively using stress-based tomography. Yao, Chang [7] presented Hough-transform method to parameterize discrete and non-Gaussian fracture properties, thus tuning the fracture network, and further proposed an integrated approach based on truncated Gaussian field to capture the spatial distribution of fractures [51]. Ma, Zhang [10] developed a multiscale parameterization method on fractured reservoir inversion based on data-driven EA. Zhang, Zhang [52] integrated deep sparse auto-encoder and ES with multiple data assimilation for naturally fractured reservoirs. Ringel, Jalali [11] introduced a flexible three-dimensional DFN inversion structure by hydraulic tomography with MCMC method. The main advantage of the structure is that the fracture number is adjustable. Zhou, Roubinet [8] presented DNN-based inversion approach to infer fracture density and fractal dimension parameters with particle-based heat-transfer model. When dealing with non-Gaussian or discrete fracture parameters, these studies can hardly generate updated ensembles satisfying prior constraint information.

This research aims to propose an efficient and effective inverse modeling framework for fracture network characterization. The hierarchical parameterization method is introduced: for small number of large fractures, each fracture is characterized by geometric parameters, i.e., fracture length, azimuth and coordination of fracture centers; for dense small fractures, fracture density and fractal dimension are utilized to characterize the fracture networks. Subsequently, variational auto-encoder and generative adversarial network (VAE-GAN) are combined to capture the strongly non-linear distribution of the parameters of complex fracture networks. VAE-GAN combines the advantage of VAE that can produce an encoder mapping data into latent space and the advantage of GAN that can achieve high-quality generative model. Moreover, ES is used under Bayesian framework based on the observing data from hydraulic tomography to infer the geometry of the fractured networks and other statistic parameters. After reducing the uncertainty of the distribution of fracture network with observing data, further engineering work and decision making can be highly facilitated.

The novelty of this study includes: (a) the hierarchical parameterization method is used to infer fracture distribution efficiently; (b) VAE-GAN is employed as generative model fusing the GAN objective with prior constraint information to map the high-dimensional discrete and non-linear parameter field into low-dimensional continuous parameter field, making it easier to generate fracture parameters under prior constraint information.

## 2. Problem statement

The geometry and stochastic parameters of the fractured networks are difficult to be obtained via direct measurements, and need to be mostly inferred from solving inverse problems with observing data. Commonly used investigation techniques for subsurface systems are hydraulic tomography [53], stress-based tomography [50], tracer test [9, 53], thermal experiments [8] and geophysical signals [5, 54]. Hydraulic tomography uses multilevel pumping tests to collect time-series observations of pressure signals by injecting and pumping fluids. According to the Bayesian theorem, the distribution of subsurface fractures can be estimated by maximizing the posterior probability distribution using the collected observation data. In this work, hydraulic tomography is employed as the experiment method to acquire measurement data.

The forward problem involves predicting the response with certain parameters by performing simulation. Thus, forward simulation for calculating fluid dynamics in fracture network is significant for estimating the fracture parameters. The true data $\mathbf{d}$ for the model $\mathbf{m}$ is obtained by running high fidelity simulation:

$$g(\mathbf{m}) = \mathbf{d} \tag{1}$$

where $g(\cdot)$ is the forward simulation process of fluid flow in fractured porous medium. After collecting the observing data $\mathbf{d}_{obs}$ using hydraulic tomography, the inverse problem for the subsurface fracture system can be modeled as follows:

$$\mathbf{d}_{obs} = g(\mathbf{m}) + \boldsymbol{\varepsilon} \tag{2}$$

where $\boldsymbol{\varepsilon}$ is the observation noise vector. The objective of this work is to infer true parameters of the model $\mathbf{m}_{true}$ based on the observing data with noise. Nevertheless, due to the limitation of observing data, the solution of the inverse problem is generally not unique. Thus, from probabilistic view, the inverse problem can be transformed into estimation of posterior probability density function $p(\mathbf{m}|\mathbf{d}_{obs})$. According to the Bayesian theorem, the posterior probability density function can be calculated as:

$$p(\mathbf{m}|\mathbf{d}_{obs}) = \frac{p(\mathbf{m})p(\mathbf{d}_{obs}|\mathbf{m})}{\int p(\mathbf{m})p(\mathbf{d}_{obs}|\mathbf{m})d\mathbf{m}} \propto p(\mathbf{m})L(\mathbf{d}_{obs}|\mathbf{m}) \qquad (3)$$

where $p(\mathbf{m})$ is the prior probability, and $L(\mathbf{d}_{obs}|\mathbf{m})$ is the likelihood function. Assuming the measurement errors and prior parameters obey Gaussian distribution, the objective function can be summarized as [55]:

$$O(\mathbf{m}) = \frac{1}{2}(\mathbf{d}_{obs} - g(\mathbf{m}))^{\mathrm{T}} C_D^{-1}(\mathbf{d}_{obs} - g(\mathbf{m})) + \frac{1}{2}(\mathbf{m} - \mathbf{m}_{pr})^{\mathrm{T}} C_M^{-1}(\mathbf{m} - \mathbf{m}_{pr}) \qquad (4)$$

where $C_D$ is the covariance of the measurement noise $\boldsymbol{\varepsilon}$, $\mathbf{m}_{pr}$ is the prior realizations, and $C_M$ is the covariance of the parameter vector $\mathbf{m}$.

## 3. Inverse modeling of fracture network with hydraulic tomography

In this section, the content is divided into five parts: parameterization of the fracture field, forward simulation of fluid flow in fracture networks, ensemble smoother for inverse modeling, deep generative model for generating fracture field, and workflow of the algorithm framework.

### 3.1. Parameterization of the fracture field

To estimate the parameters of the fracture network, parameterization technique is a necessary step to characterize the complex fracture field effectively. Existing parameterization methods can be classified into transform-based methods [56], equivalent methods [57], geometrical methods of discrete fractures [52] and fractal methods [8]. The geometry of a 2D fracture can be parameterized in two different ways, namely coordinate parameterization and angle/radius parameterization (**Fig. 1**), with the former to be easier to determine and has less constraints [58]. The fractal method is advantageous in dense and small fracture parameterization. In this work, multiscale-parameterization method integrating coordinate and fractal parameterization is employed to infer large and small fracture fields hierarchically and construct the complex fracture networks. The large fractures are characterized with the fracture length, orientation and coordinate of fracture center, while the small fractures are generated by fractal theories in a statistical way. Consequently, the high-dimensional parameter can be further estimated with inverse modeling using observing data.

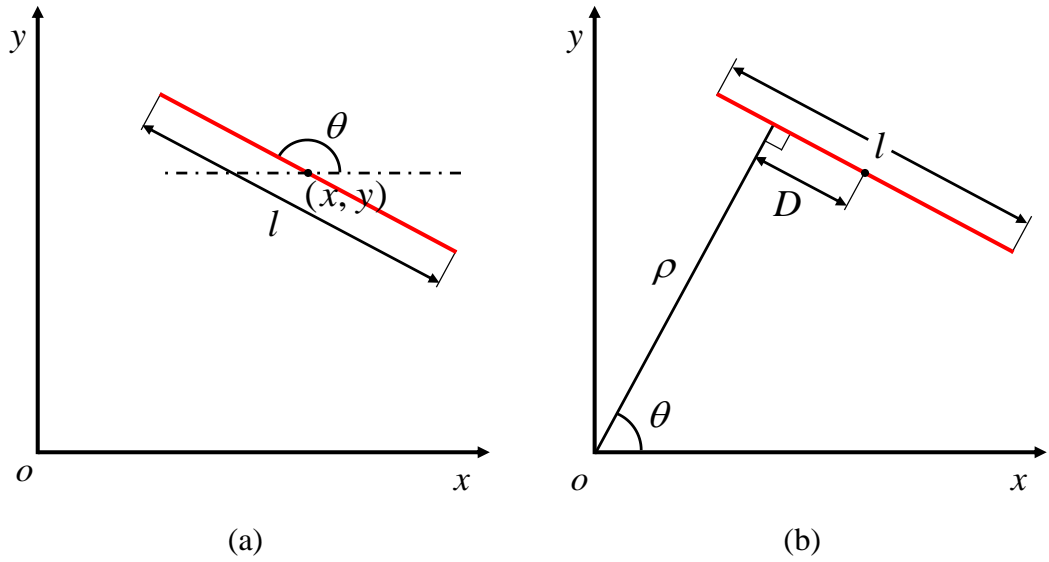

(a)          (b)

**Fig. 1.** Two different geometrical parameterization ways for a 2D fracture [7, 10].

For large fractures in 2D space, the parameters are characterized as

$$\mathbf{m}_l = \{\mathbf{x}_i, \mathbf{y}_i, \mathbf{\theta}_i, \mathbf{l}_i\}_{i=1,\dots,N_l} \tag{5}$$

where $\mathbf{m}_l$ is the parameter vector, $(\mathbf{x}_i, \mathbf{y}_i)$ is the midpoint coordinate of the $i^{th}$ large fracture, $\mathbf{\theta}_i$ is the orientation of the $i^{th}$ large fracture, and $\mathbf{l}_i$ is the length of the $i^{th}$ large fracture.

For small fractures, fractal theory is adopted to generate 2D fractal discrete fracture network. $N(l)$ denotes the number of fractures with length greater than $l$, and the fracture length $l$ follows a power law expressed as:

$$N(l) = C(l_{max}/l)^{D_f} \tag{6}$$

where $C$ is the fracture density, $D_f$ is the fractal dimension, and $l_{max}$ is the maximum fracture length. Thus, the total number of fractures $N_{total}$ is

$$N_{total} = N(l_{min}) = C(l_{max}/l_{min})^{D_f} \tag{7}$$

where $l_{min}$ is the minimum fracture length. **Fig. 2** shows the relationship between cumulative fracture number and fracture length obeying power-law distributing under different fractal dimensions and fracture densities.

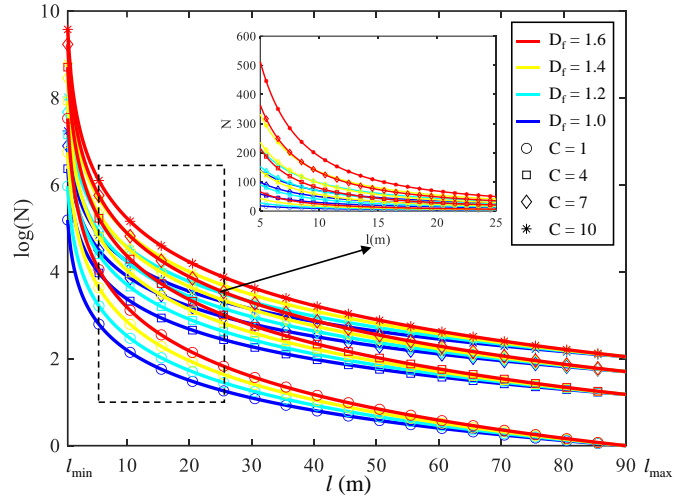

**Fig. 2.** Cumulative fracture number vs. fracture length obeying power-law distributing under different fractal dimensions and fracture densities.

Besides these parameters, the aperture of the fracture is also significant for fractured flow. The aperture of fractures can be set as constant, which is proportional to the fracture lengths, uniform, and in lognormal distribution. To simplify the problem, the aperture of fractures is assumed to be proportional to the fracture length:

$$d_f = \alpha l \tag{8}$$

where $d_f$ is the aperture (m) of the fracture, $\alpha$ is a constant, and $l$ denotes the fracture length (m). According to the cubic law [59], the permeability ($k_f$) of the fracture with aperture $d_f$ can be calculated by:

$$k_f = \frac{d_f^2}{12} \tag{9}$$

### 3.2. Forward simulation of fluid flow in fracture networks

Using numerical simulation to calculate the flow dynamics of the fracture system provides a powerful way to better understand the unknown geometry and statistic parameters of the fracture networks. In this study, fluid flow is considered to be Darcy's flow and the Darcy velocity $u$ can be calculated by:

$$u = -\frac{k}{\mu}(\nabla p + \rho g \nabla D) \tag{10}$$

where $k$ is the permeability of porous medium ($m^2$), $\mu$ is the viscosity of fluid ($Pa \cdot s$), $\nabla p$ is the pressure gradient ($Pa/m$), $\rho$ is the density of fluid ($kg/m^3$), g is the constant of gravitational acceleration ($m/s^2$), and $\nabla D$ is a unit vector along the direction of gravity. According to the mass conservation principle, the governing equation of fluid flow is:

$$\frac{\partial}{\partial t}(\rho \phi) + \nabla(\rho u) = Q_m \tag{11}$$

where $t$ is the time (s), $\phi$ is the porosity of porous medium, and $Q_m$ is the mass source/sink term ($kg/(m^3 \cdot s)$). Fluid flow in fractures is considered to obey Darcy's law at the physics interface:

$$q_f = -\frac{k_f}{\mu} d_f (\nabla_T p + \rho g \nabla_T D) \tag{12}$$

where $q_f$ is the volumetric flow rate per unit length of the fracture ($m^2/s$), $\nabla_T$ is the gradient in the tangential plane of fracture, D denotes the vertical coordinate (m). Thus, the mean velocity of fluid $u_f$ ($m/s$) in the fracture can be obtained:

$$u_f = q_f / d_f \tag{13}$$

Integrating cubic law, model properties and governing equation, the pressure control equation can be derived as follows:

$$d_f \frac{\partial}{\partial t}(\rho \phi_f) + \nabla_T (\rho q_f) = d_f Q_m \tag{14}$$

where $\phi_f$ is the porosity of fracture. In this study, the physical system governed by partial differential equation is calculated by finite element method using COMSOL Multiphysics.

### 3.3. Ensemble smoother for inverse modeling

Dealing with simulation-based inverse problems generally involves a large amount of forward simulation evaluations. Therefore, how to estimate the geometry of the fractured networks and quantify corresponding uncertainties are indispensable for further optimization and decision making of fractured aquifers. MCMC and EAs are popular to estimate the posterior distribution of parameter fields to quantify the uncertainty of subsurface model. However, these methods need a large amount of time-consuming forward model execution to obtain robust optimization. ES is adopted in this work on solving the inverse problems. A group of realizations are generated based on prior information. After using numerical simulation to calculate fluid dynamics, the ensemble of the model parameters is updated as

$$\mathbf{m}_i^{(j+1)} = \mathbf{m}_i^{(j)} + C_{MY}^{(j)} (C_{YY}^{(j)} + C_D)^{-1} [\mathbf{d}_{obs} + \varepsilon_i - g(\mathbf{m}_i^{(j)})] \tag{15}$$

where $C_{MY}^{(j)}$ is the covariance between model parameters $\mathbf{m}^{(j)}$ for the j$^{th}$ ensemble and the corresponding predictions, and $C_{YY}^{(j)}$ is the auto-covariance of the predictions for the j$^{th}$ ensemble. Only single update of the ensemble for highly nonlinear problems is not sufficient. Thus, we adopt multiple data assimilation to solve the fracture inversion problem [60]. The updating formula is further modified by introducing inflation factor:

$$\mathbf{m}_i^{(j+1)} = \mathbf{m}_i^{(j)} + C_{MY}^{(j)} [C_{YY}^{(j)} + (\alpha_t)^2 C_D]^{-1} [\mathbf{d}_{obs} + \alpha_t \varepsilon_i - g(\mathbf{m}_i^{(j)})] \tag{16}$$

where $\alpha_t$ is the inflation factor which can be set as $\alpha_t = \sqrt{N_{iter}}$ [28]. **Fig. 3** presents the overview of the ensemble smoother algorithm. The detailed procedure of ES in dealing with DFN inversion problems is: Initialize discrete fracture network realizations with prior information; generate initial ensemble; perform forward simulation to calculate the prediction of fluid dynamics (blue curve) for each realization in the ensemble and compare with observing data (red curve); calculate the root mean square error (RMSE) and update the covariance matrix $C_{MY}$, if the stopping criterion is satisfied, then stop, otherwise the loop continues. The ensemble is finally converted to a probability map to quantify the uncertainty of the fracture distributions.

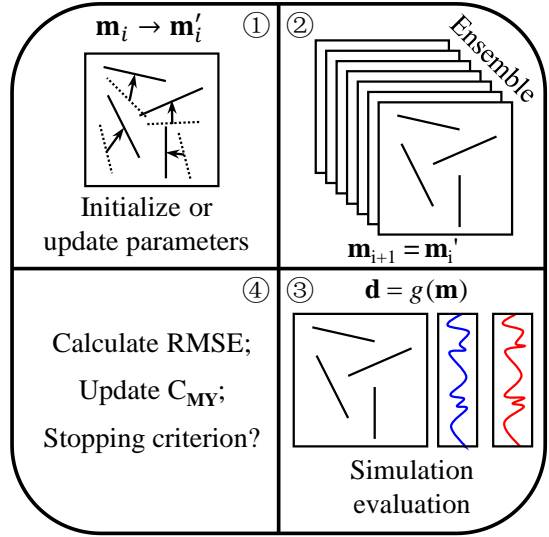

**Fig. 3.** Overview of the ensemble smoother algorithm.

### 3.4. Deep generative model

In this study, DNN is employed to explore the underlying relationship between the input parameters and the lower-dimensional latent variables with hierarchy of hidden layers to efficiently estimate the distribution of parameters. Then the ES method can sample the posterior distribution of the model parameters in a consistent and coherent manner.

### 3.4.1. Variational auto-encoder

Variational auto-encoder (VAE), a generative model, is able to transform the high-dimensional parameters into low-dimensional latent variables with encoder and transform the latent representation back to the high-dimensional parameters with decoder, respectively [61]. VAE consists of encoder and decoder, which can be mathematically expressed as:

$$\mathbf{z} \sim \text{Enc}(\mathbf{x}) = q(\mathbf{z}|\mathbf{x}), \quad \hat{\mathbf{x}} \sim \text{Dec}(\mathbf{z}) = p(\mathbf{x}|\mathbf{z}) \tag{17}$$

VAE imposes a prior for the latent distribution $p(\mathbf{z})$ as the regularization term, generally choosing normal distribution $\mathbf{z} \sim \mathcal{N}(\mathbf{0}, \mathbf{I})$. The objective of VAE is to minimize the reconstruction error and to make latent variables close to normal distribution. Thus, the loss function of VAE $\mathcal{L}_{\text{VAE}}$ is minus the sum of the expected log likelihood function (i.e., the reconstruction error) and prior regularization:

$$\mathcal{L}_{\text{VAE}} = -\mathbb{E}_{q(\mathbf{z}|\mathbf{x})}[\log \frac{p(\mathbf{x}|\mathbf{z})p(\mathbf{z})}{q(\mathbf{z}|\mathbf{x})}] = \mathcal{L}_{\text{like}}^{\text{pixel}} + \mathcal{L}_{\text{prior}} \tag{18}$$

with

$$\mathcal{L}_{\text{like}}^{\text{pixel}} = -\mathbb{E}_{q(\mathbf{z}|\mathbf{x})}[\log p(\mathbf{x}|\mathbf{z})] \tag{19}$$

$$\mathcal{L}_{\text{prior}} = D_{\text{KL}}(q(\mathbf{z}|\mathbf{x}) \| p(\mathbf{z})) \tag{20}$$

where $\mathcal{L}_{\text{like}}^{\text{pixel}}$ denotes the reconstruction error, $\mathcal{L}_{\text{prior}}$ denotes the prior regularization term, and $D_{\text{KL}}$ represents the Kullback-Leibler divergence, which is a statistical distance measuring the distance between the prior regularization $p(\mathbf{z})$ and the distribution $q(\mathbf{z}|\mathbf{x})$.

### 3.4.2. Generative adversarial network

Generative adversarial network (GAN) is also a generative model consisting of a generator G and a discriminator D [62]. The generator learns a mapping from low-dimensional latent variables to true sample, while the discriminator learns to distinguish generated samples from true samples. The framework corresponds to a minimax two-player game. The final result is that the generative model is able to capture the distribution of data and the discriminator is able to discriminate generated samples and true samples [63]. Thus the objective is to maximize/minimize the following cross entropy [64]:

$$\min_G \max_D \mathcal{L}_{\text{GAN}}(G, D) = \mathbb{E}_{\mathbf{x} \sim p(\mathbf{x})}[\log(D(\mathbf{x}))] + \mathbb{E}_{\mathbf{z} \sim p(\mathbf{z})} \log[1 - D(G(\mathbf{z}))] \tag{21}$$

### 3.5. Workflow of the deep learning based inversion framework

In this study, a variational generative adversarial network combining VAE and GAN is introduced to capture the strongly non-linear distribution of the parameters of complex fracture networks. VAE-GAN is able to map the high-dimensional discrete and non-linear parameter field into low-dimensional continuous parameter field. VAE-GAN combines the advantage of VAE that can produce an encoder mapping data into latent space and the advantage of GAN that can achieve high-quality generative model. Subsequently, ES is employed under Bayesian framework based on the observing data from hydraulic tomography to infer the geometry of the fractured networks and other statistic parameters. The updated ensemble is finally converted to a probability map to quantify the uncertainty of the fracture distributions. After reducing the uncertainty of the distribution of fracture network with the proposed deep learning based inversion framework, further engineering work and decision making can be performed.

The variational generative adversarial network consists of three parts: encoder $\text{Enc}$, decoder or generator $\text{Dec\_G}$, and discriminator D. VAE-GAN incorporates the decoder of VAE and the generator of GAN into one by sharing parameters and training jointly. By combining VAE with GAN, the feature representation learned in the discriminator of GAN can be used as the basis for the VAE reconstruction target. The structure of the three networks used in this work are multilayer perceptron. The activation function of the hidden layers is tanh, and the activation function of the output layers is sigmoid function to ensure the output range is in [0, 1].

To generate the fracture network that can satisfy the prior constraint information, we introduce a loss function for generator of GAN as follows:

$$\mathcal{L}_{\text{constraint}}(\text{Dec\_G}) = c(\text{Dec\_G}(\mathbf{z})) \tag{22}$$

where $c(\cdot) \geq 0$ is the prior constraint function. Thus, the final objective of generator is:

$$\text{Dec\_G} = \arg\min_{\text{Dec\_G}} \max_{D} \mathcal{L}_{\text{GAN}}(\text{Dec\_G,D}) + \lambda \mathcal{L}_{\text{constraint}}(\text{Dec\_G}) \tag{23}$$

To tune the weight of each objective, the weight factors are introduced. During the training process, the parameters of encoder $\theta_{\text{Enc}}$, decoder or generator $\theta_{\text{Dec\_G}}$ and discriminator $\theta_{\text{D}}$ are updated as:

$$\theta_{\text{Enc}} \pm -\nabla_{\theta_{\text{Enc}}} (\lambda_1 \mathcal{L}_{\text{prior}} + \lambda_2 \mathcal{L}_{\text{like}}^{\text{pixel}}) \qquad (24)$$

$$\theta_{\text{Dec\_G}} \pm -\nabla_{\theta_{\text{Dec\_G}}} [\lambda_2 \mathcal{L}_{\text{like}}^{\text{pixel}} + \lambda_3 \mathcal{L}_{\text{GAN}}(\text{Dec\_G},\text{D}) + \lambda_4 \mathcal{L}_{\text{constraint}}(\text{Dec\_G})] \qquad (25)$$

$$\theta_{\text{D}} \pm -\nabla_{\theta_{\text{D}}} [\lambda_3 \mathcal{L}_{\text{GAN}}(\text{Dec\_G},\text{D}) + \lambda_4 \mathcal{L}_{\text{constraint}}(\text{Dec\_G})] \qquad (26)$$

After fusing complex prior geological information, the networks not only learn the mapping from input sample to output sample, but also learn a loss function to train this mapping. The framework of the VAE-GAN for generating fracture network and the flow diagram of the stochastic deep learning inversion is presented in **Fig. 4**. The pseudo-code for training VAE-GAN model is presented in **Algorithm 1**.

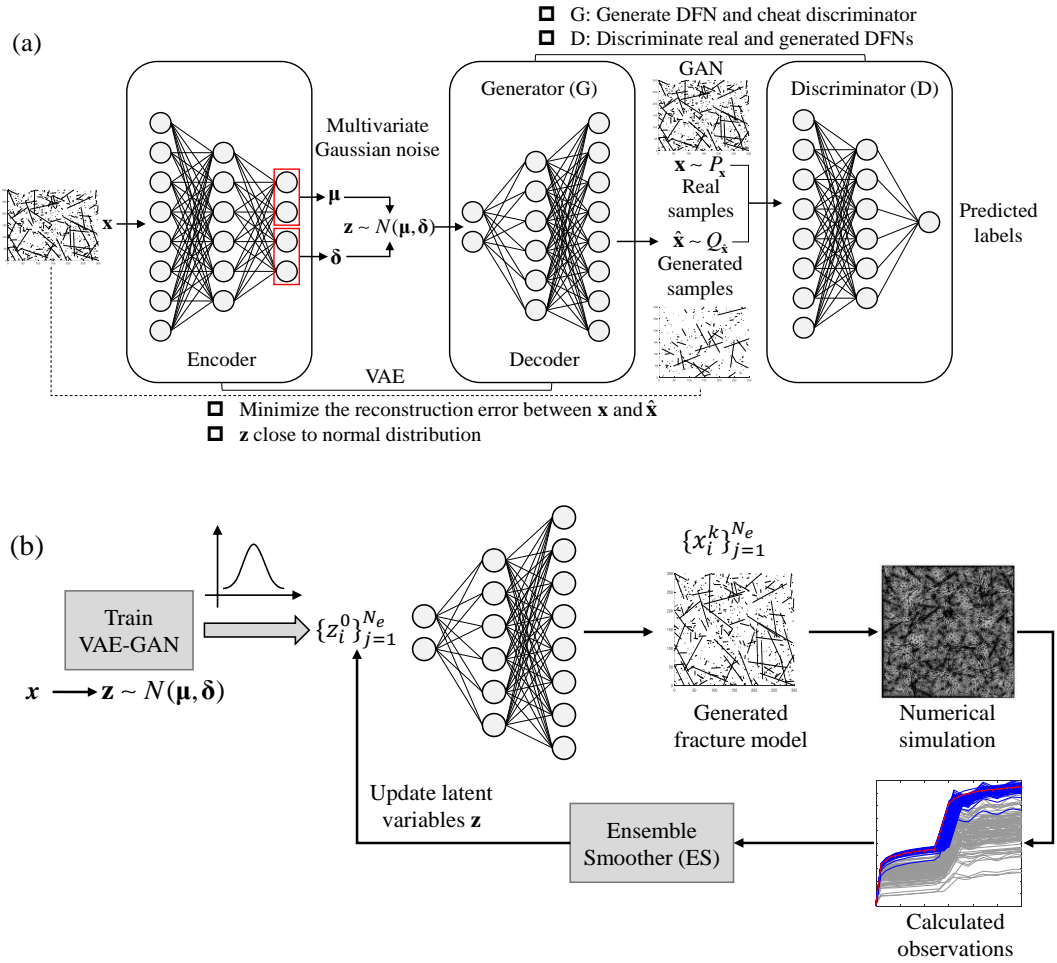

**Fig. 4.** (a) The framework of the generative adversarial network for generating fracture networks. (b) Flow diagram of the stochastic deep learning inversion.

**Algorithm 1** Training VAE-GAN algorithm

**Input**: $\theta_{\text{Enc}}, \theta_{\text{Dec\_G}}, \theta_{\text{D}} \leftarrow$ initialize encoder, decoder or generator, and discriminator network parameters, weight factor $\lambda_1, \lambda_2, \lambda_3$ and $\lambda_4$

**While** stopping criterion has not met

  $\mathbf{x} \leftarrow$ random training mini-batch samples;

  $\mathbf{z} \leftarrow \text{Enc}(\mathbf{x})$;

  $\mathcal{L}_{\text{prior}} \leftarrow D_{\text{KL}}(q(\mathbf{z}|\mathbf{x}) \| p(\mathbf{z}))$;

  $\hat{\mathbf{x}} \leftarrow \text{Dec\_G}(\mathbf{z})$;

  $\mathcal{L}_{\text{like}}^{\text{pixel}} \leftarrow -\mathbb{E}_{q(\mathbf{z}|\mathbf{x})}[\log p(\mathbf{x}|\mathbf{z})]$;

  $\mathbf{z}_p \leftarrow$ samples from prior distribution $\mathcal{N}(\mathbf{0}, \mathbf{I})$;

  $\hat{\mathbf{x}}_p \leftarrow \text{Dec\_G}(\mathbf{z}_p)$;

  $\mathcal{L}_{\text{GAN}}(\text{Dec\_G}, \text{D}) = \mathbb{E}_{\mathbf{x} \sim p(\mathbf{x})}[\log(\text{D}(\mathbf{x}))] + \mathbb{E}_{\mathbf{z}_p \sim p(\mathbf{z}_p)} \log[1 - \text{D}(\hat{\mathbf{x}}_p)]$;

  $\mathcal{L}_{\text{constraint}}(\text{Dec\_G}) = c(\hat{\mathbf{x}}_p)$;

  // Update parameters according to gradients;

  $\theta_{\text{Enc}} \xleftarrow{\pm} -\nabla_{\theta_{\text{Enc}}}(\lambda_1 \mathcal{L}_{\text{prior}} + \lambda_2 \mathcal{L}_{\text{like}}^{\text{pixel}})$;

  $\theta_{\text{Dec\_G}} \xleftarrow{\pm} -\nabla_{\theta_{\text{Dec\_G}}}[\lambda_2 \mathcal{L}_{\text{like}}^{\text{pixel}} + \lambda_3 \mathcal{L}_{\text{GAN}}(\text{Dec\_G}, \text{D}) + \lambda_4 \mathcal{L}_{\text{constraint}}(\text{Dec\_G})]$;

  $\theta_{\text{D}} \xleftarrow{\pm} -\nabla_{\theta_{\text{D}}}[\lambda_3 \mathcal{L}_{\text{GAN}}(\text{Dec\_G}, \text{D}) + \lambda_4 \mathcal{L}_{\text{constraint}}(\text{Dec\_G})]$;

**End while**

**Output**: decoder or generator $\text{Dec\_G}$

The performance of commonly-adopted ES is not well when dealing with complex non-Gaussian parameter distribution. The proposed deep learning based inversion framework is able to capture the strongly non-linear distribution of the parameters of complex fracture network and transform the original space into normally distributed latent space. By employing deep generative model, the random variables $\mathbf{z}$ with normal distribution $\mathcal{N}(\mathbf{0}, \mathbf{I})$ can transform to high-dimensional model parameters with the deep generative model. Therefore, the ES for parameter inversion of fracture network with Eq. 16 can be rewritten as follows:

$$\mathbf{z}_i^{(j+1)} = \mathbf{z}_i^{(j)} + C_{\mathbf{zY}}^{(j)}(C_{\mathbf{YY}}^{(j)} + (\alpha_t)^2 C_D)^{-1}[\mathbf{d}_{\text{obs}} + \alpha_t \varepsilon_i - g(\text{Dec\_G}(\mathbf{z}_i^{(j)}))] \qquad (27)$$

The pseudo-code of the deep learning based ES for fracture inversion is presented in **Algorithm 2**.

---
**Algorithm 2** Deep learning based ES for fracture inversion
---
**Input**: deep generative model $\text{Dec\_G}$.

Train the generative model VAE-GAN;

Initialize DFN realizations and generate initial ensemble $\mathbf{z} \sim \mathcal{N}(\mathbf{0}, \mathbf{I})$;

**While** stopping criterion has not met

    Update the ensemble $\mathbf{z}$;

    Calculate the high-dimensional DFN realizations $\mathbf{m} = \text{Dec\_G}(\mathbf{z})$;

    Perform forward simulation to calculate the prediction of fluid dynamics $\mathbf{Y}$;

    Calculate the root mean square error and update the covariance matrix $C_{\mathbf{zY}}$;

**End while**

Generate a probability map using the updated ensemble to quantify the uncertainty of the fracture distribution;

**Output**: Database D

---

## 4. Case studies

To test the performance of the proposed deep learning based inversion framework on estimating fracture network parameters, two 2D synthetic numerical examples are presented. The first case has a designated number of large fractures in the prior information, and the second case presents a more complex fracture network containing both large and small fractures.

### 4.1. Synthetic hydraulic tomography experiment

The synthetic hydraulic tomography experiment employs multilevel pumping tests to collect time-series observations of pressure signals by injecting and producing fluids. The goal is to infer the fracture geometry and statistic parameters of fracture network from the observing pressure curve. According to the Bayesian theorem, the distribution of subsurface fractures can be estimated by maximizing the posterior probability distribution. It is worth noting that the observing data can also be obtained by stress-based tomography, tracer test, thermal experiments, geophysical signals, and dynamic data of hydrocarbon exploitation, etc [9, 50].

**4.2. Case 1: Fracture network with azimuth along one direction**

In this case, the fracture field contains 15 fractures with azimuth along one direction. There are totally 25 drilling wells for the aquifer within a $100\times100$ m region. The model in case 1 is a 2D horizontal model, ignoring the effect of gravity on fluid flow. The true fracture distribution and parameter settings of the fracture network model for case 1 is presented in **Fig. 5** and **Table 1**, respectively. The prior information for case 1 is: the model contains 15 fractures with mean trend $120°$ and variance 15; the fracture lengths range from 20 to 60 meters. Since the number of the fractures is known and small, each fracture is characterized by the length, azimuth and coordination of the fracture center. Thus, the total number of variables to be estimated is 60. To estimate the locations of the fractures, hydraulic tomography experiment is performed with 1 injection well, 4 production wells, and 20 monitoring wells. For the hydraulic tomography experiment setting, the injection well $I_1$ is set at constant injection rate of $10^{-2}\,\mathrm{m^3/s}\,(10\text{ kg/s})$ for the first 3000 s and $8\times10^{-3}\,\mathrm{m^3/s}$ for the late 3000 s. The pumping rates of the production wells $P_1$ and $P_4$ are set to be $1\times10^{-3}\,\mathrm{m^3/s}\,(1\text{ kg/s})$ for the first 3000 s and $1.5\times10^{-3}\,\mathrm{m^3/s}$ (1.5 kg/s) for the late 3000 s, while the pumping rates of the production wells $P_2$ and $P_3$ are set to be $2\times10^{-3}\,\mathrm{m^3/s}\,(2\text{ kg/s})$ for the first 3000 s and $1\times10^{-3}\,\mathrm{m^3/s}\,(1\text{ kg/s})$ for the late 3000 s. The standard deviation of the noise for the observations is set to 1% of the range for pressure predictions to generate the synthetic measurement.

**Table 1** Parameter settings of the fracture network model

| Parameter | Value |
| --- | --- |
| Model thick | 5 m |
| Permeability of matrix | $10^{-11}$ m$^2$ |
| Porosity of matrix | 0.1 |
| Permeability of fractures | $10^{-7}$ m$^2$ |
| Porosity of fractures | 0.3 |
| Specific storage | $1.2 \times 10^{-8} \, \text{Pa}^{-1}$ |
| Mean trend | 120 |
| Trend variance | 15 |
| Fracture length range | [20, 60] |

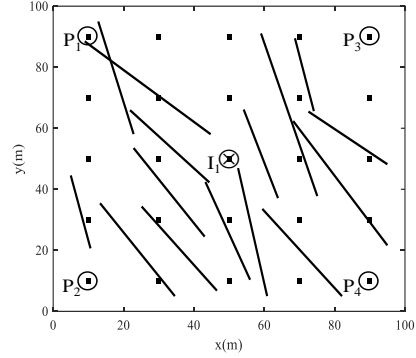

**Fig. 5.** True fracture distribution of the fracture network model for case 1.

To generate conditional fracture parameters with the generative model, 10000 fracture fields satisfying prior information and constraints are employed as training samples to learn the corresponding parameter distributions. 2000 fracture fields are used for validation. The dimension of the latent space is set to 40. Based on prior information, observation data and well-trained generative model, ES is applied with $N_e = 100$ at 25 wells for $N_{iter}$ data assimilation iterations.

To illustrate the change of fracture parameters in low-dimensional latent space, two low-dimensional latent parameter vectors $\mathbf{z_1}$ and $\mathbf{z_2}$ are adopted to visualize the variation in the latent space by linearly interpolating using different $\gamma$ values, and then reparametrized with the generator as expressed by following equation:

$$\text{Dec\_G}(\mathbf{z}) = \text{Dec\_G}[\mathbf{z_1} + \gamma(\mathbf{z_2} - \mathbf{z_1})] \tag{28}$$

**Fig. 6** presents the gradual change of fracture distributions from $\mathbf{z_1}$ to $\mathbf{z_2}$ after re-parameterization with different $\gamma$ value. From (a) to (j), it presents the reconstructed fracture network with γ ranges from 0 to 1 in Eq. 28. **Fig. 7** presents the iterations of the fracture inversion process for case 1 using probabilistic map of fracture ensembles and randomly selected fracture distribution realizations. Before the inversion, the fractures appear blurry in the probabilistic map. During the inversion process, the probabilistic map of fracture ensembles gradually presents the possible azimuth and location of each fracture. **Figs.** 7j, 7o, 7t, and 7y present the updated ensemble of four randomly selected realizations. In comparison with true fracture distribution in **Fig. 5**, the final updated realizations capture the most of fracture locations and azimuths correctly.

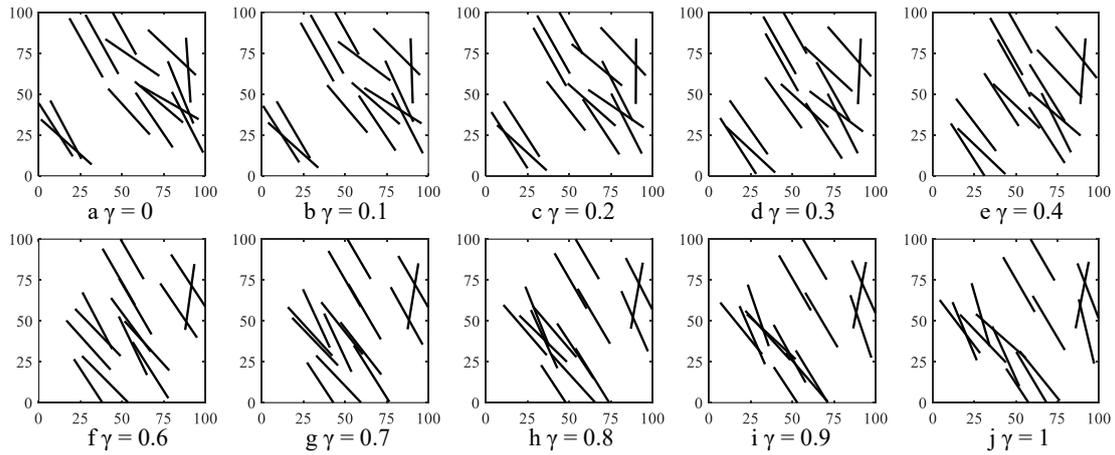

**Fig. 6.** Latent space visualization between $z_1$ and $z_2$ using generative model for case 1.

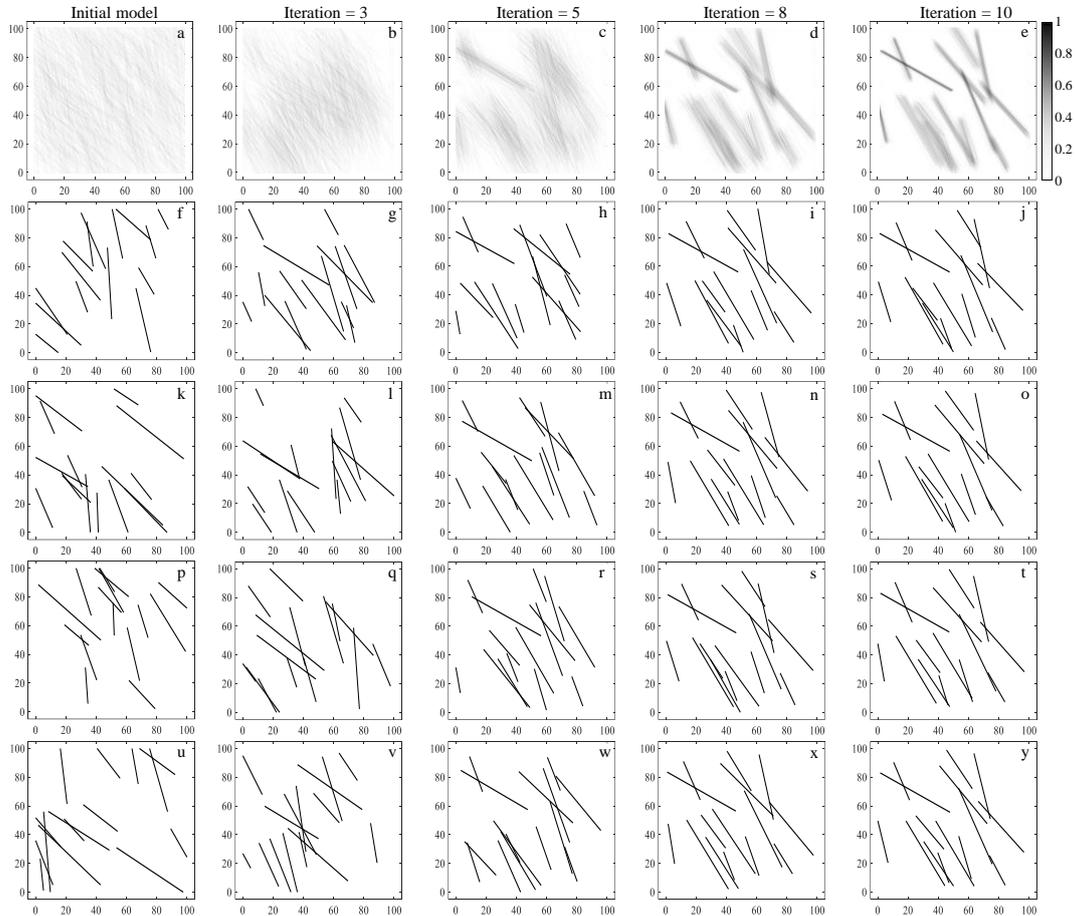

**Fig. 7.** The iterations of the fracture inversion process for case 1. The rows are the fracture inversion process of (a-e) probabilistic map of fracture ensembles, and (f-y) randomly selected fracture distribution realizations, respectively. The columns are the initial model, after 3, 5, 8, and 10 iterations during data assimilation process.

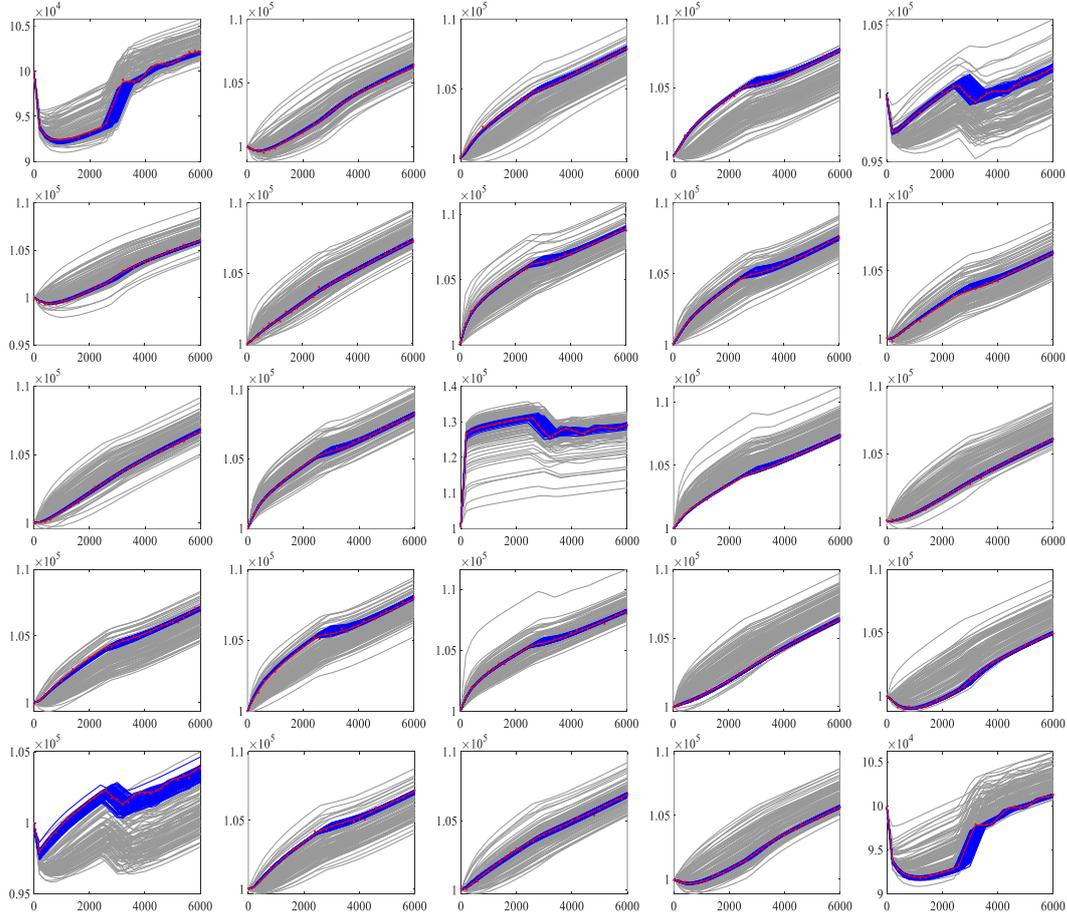

**Fig. 8.** The matching results of observing pressure for case 1. The red curves are the prediction of true fracture field, and the red points are the observing data with noise. The gray and blue curves are the predictions of initial and updated ensembles.

**Fig. 8** summarizes the matching results of observing pressure of piezometers for the case. Despite the complexity of the inversion of fracture field with statistical trend constraints, the proposed algorithm shows great performance on the fracture inversion problem of fine characterization of a small number of large fractures. The updated ensemble shows significantly reduced uncertainty of pressure data matching of the fracture field. The prior distributions and posterior distributions of the 40 latent variables for the initial ensemble and the updated ensemble after data assimilation are illustrated in **Fig. 9**. The prior distributions of the latent variables are standard normal distribution. After data assimilation, the posterior distributions of the latent variables converge to a range with small variability. Using the parameters of fracture location, azimuth and length to characterize each fracture is efficient and effective for the fracture fields with a small number of large fractures. To further demonstrate the performance of the proposed algorithm, as present in **Fig. 10**, the pressure prediction for a randomly selected posterior fracture realization at 6000 s is compared with the reference model. The fracture model can get the main trend of the pressure distribution. When dealing with inverse modeling of complex fracture networks, fine characterization for each fracture is impractical due to the large number of parameters.

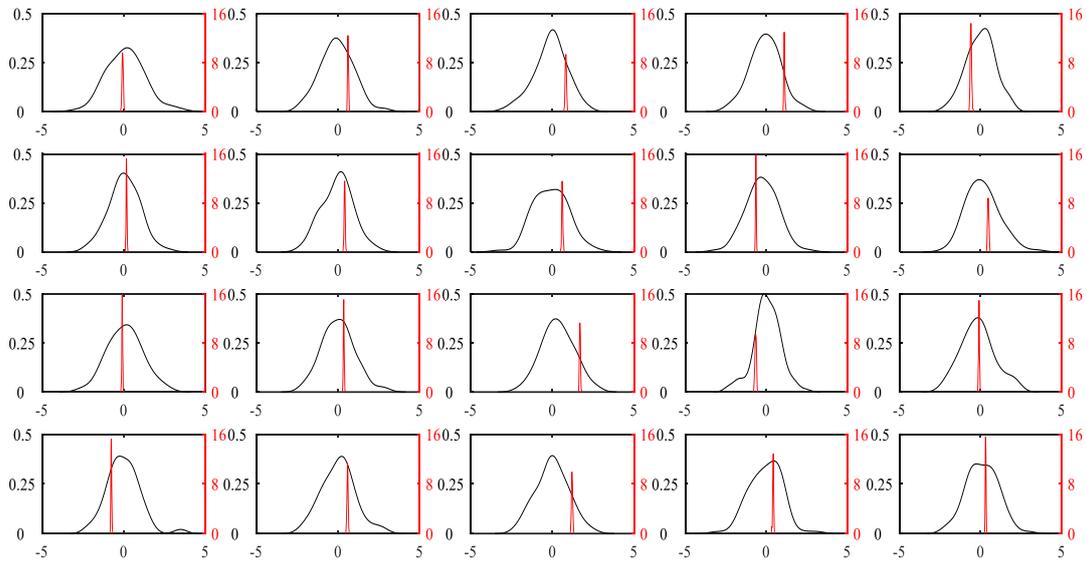

**Fig. 9.** Case 1: The prior distributions (black) of the 40 latent variables for the initial ensemble and the posterior distributions (red) of the 40 latent variables for the updated ensemble after data assimilation.

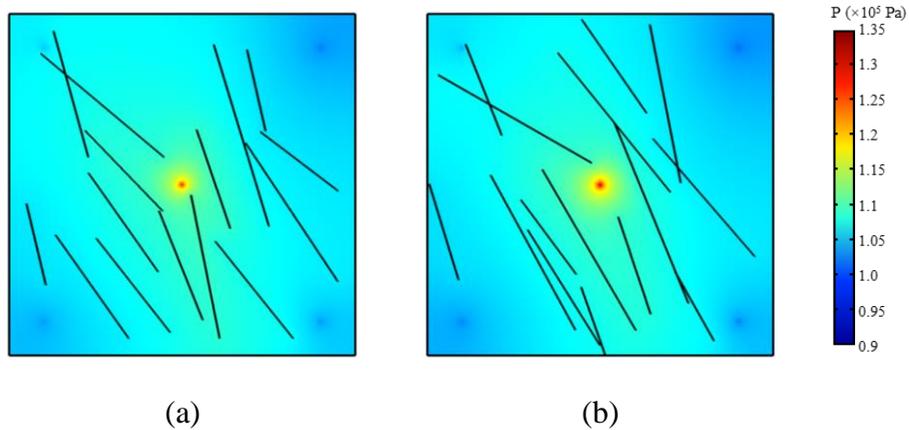

(a)          (b)

**Fig. 10.** Case 1: The pressure prediction at 6000 s for (a) the reference model, and (b) a selected posterior fracture realization.

### 4.3. Case 2: Synthetic multiscale fracture network

In case 2, a synthetic multiscale fracture network with fracture length obeying fractal theory is introduced. The model in case 2 is a 2D horizontal model, ignoring the effect of gravity on fluid flow. The flow domain is in a $100\times100$ m region, with a total of 25 drilling wells in the aquifer. The initial pressure is around $10^5$ Pa. The fracture density for the true fracture field is 2, the fractal dimension is 1.2, and the true fracture distribution is presented in **Fig. 11**, and corresponding parameter settings are illustrated in **Table 2**. The fracture network contains two main sets, with mean trend 50 for set 1 and 120 for set 2, respectively. The variance of trends for both sets 1 and 2 is 1. For this model, the locations of the top 20 large fractures, fracture density and fractal dimension are to be estimated, that is, the number of parameters to be estimated is 82.

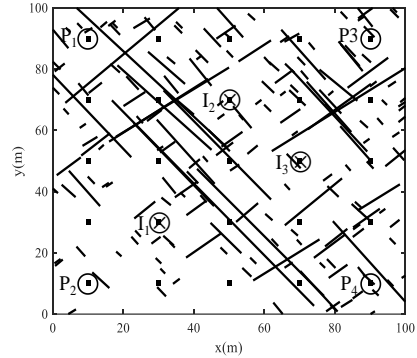

**Fig. 11.** True fracture distribution of the fracture network model for case 2.

**Table 2** Main parameter settings of the multiscale fracture network model

| Parameter | Value |
| --- | --- |
| Permeability of porous medium | $10^{-11}$ m$^2$ |
| Porosity of porous medium | 0.1 |
| Permeability of fractures | $10^{-7}$ m$^2$ |
| Porosity of fractures | 0.3 |
| Range of fracture density | [1.0, 3.0] |
| Range of fractal dimension | [1.0, 1.3] |
| Specific storage | $1.2 \times 10^{-8} \text{Pa}^{-1}$ |
| Mean trend of set 1 | 50 |
| Trend variance of set 1 | 1 |
| Mean trend of set 2 | 120 |
| Trend variance of set 2 | 1 |
| Minimum fracture length | 1 |
| Maximum fracture length | 80 |

In order to estimate the fracture density, fractal dimension, and locations of large fractures, hydraulic tomography experiment is performed in 3 injection wells, 4 pumping wells, and 18 monitoring wells. For the hydraulic tomography experiment setting, the injection wells $I_1$, $I_2$, and $I_3$ are set at constant injection rates of $2 \times 10^{-3} \text{m}^3/\text{s}$, $2 \times 10^{-3} \text{m}^3/\text{s}$, and $1.5 \times 10^{-3} \text{m}^3/\text{s}$, respectively, for the first 3000 s; set at constant injection rates of $3 \times 10^{-3} \text{m}^3/\text{s}$, $1.5 \times 10^{-3} \text{m}^3/\text{s}$, and $2 \times 10^{-3} \text{m}^3/\text{s}$, respectively, for the late 3000 s. To generate the synthetic measurement, 1 % noise of the range for pressure predictions is imposed on the synthetic observing data.

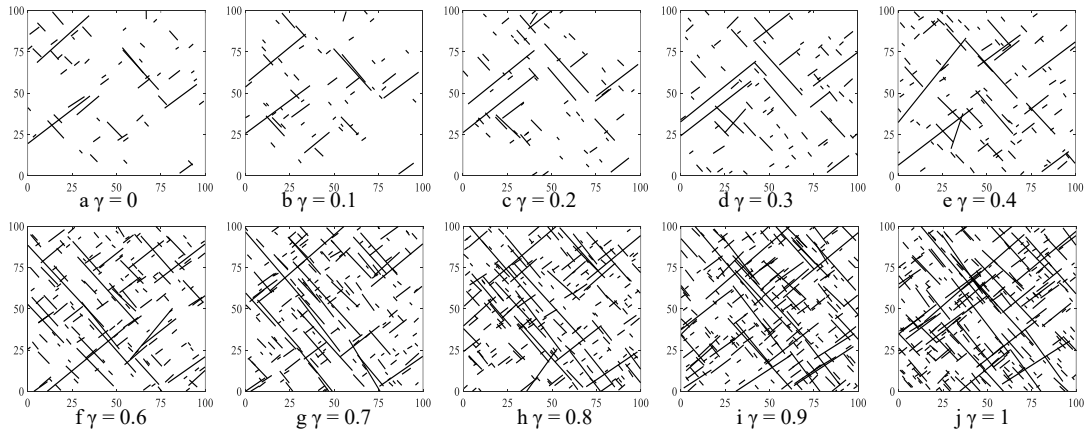

**Fig. 12.** Effect of perturbation in latent space using generative model for case 2. (a) to (j) present the reconstructed fracture network with γ ranges from 0 to 1 in Eq. 28.

In this case, 20000 fracture fields satisfying fracture length fractal theory and fracture azimuth prior constraints are employed as training samples to learn the corresponding parameter distributions. 2000 fracture fields are adopted for validation. The latent dimension is set to 40. Based on prior information, observation data and well-trained generative model, ES is applied with $N_e = 100$ at 25 wells for $N_{iter}$ data assimilation iterations. Due to some parameters of the fracture field to be estimated are statistic parameters, i.e., fracture density and fractal dimension, the fracture fields with same parameters may have different observations. Therefore, the final estimated parameters are highly uncertain with high confidence intervals. To alleviate this impact, each forward simulation result is obtained by generating multiple fracture network realizations with same parameters and calculating the average observing data value.

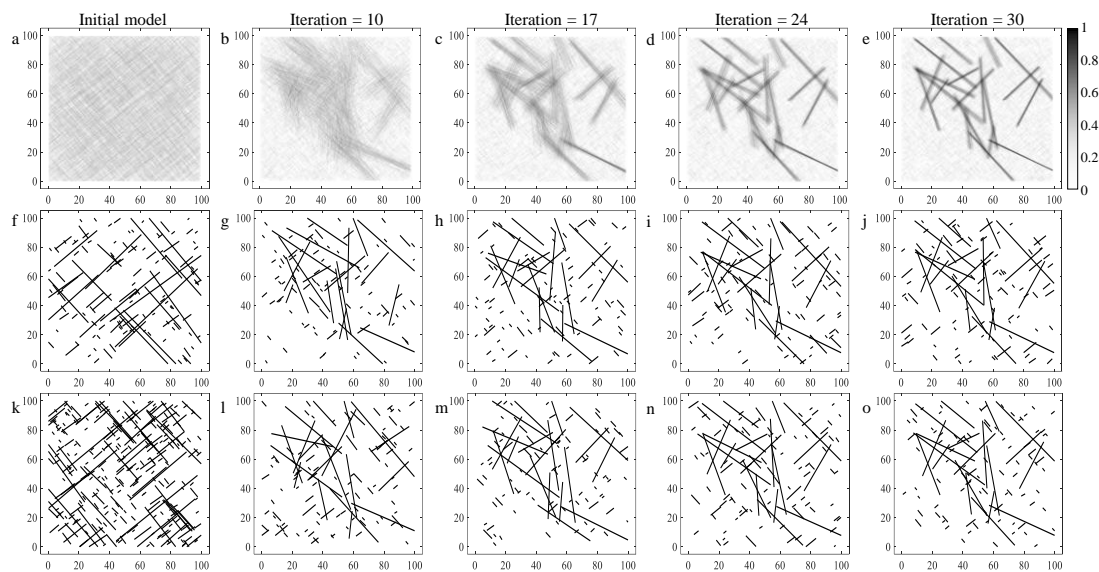

**Fig. 13.** The iterations of the fracture inversion process of ES for case 2. The rows are the fracture inversion process of (a-e) probabilistic map of fracture ensembles, and (f-o) randomly selected fracture distribution realizations, respectively. The columns are the initial model, after 10, 17, 24 and 30 iterations during data assimilation process, respectively.

To visualize the variation of latent variables of deep generative model on fracture distributions, two low-dimensional latent parameter vectors $z_1$ and $z_2$ are randomly selected. The other vectors are generated by linearly interpolating using different $\gamma$ values calculated by Eq. 28, and then reparametrized with the generator, as presented in **Fig. 12**. The density of fracture networks gradually increases with $\gamma$, and the prior constraints have been satisfied well.

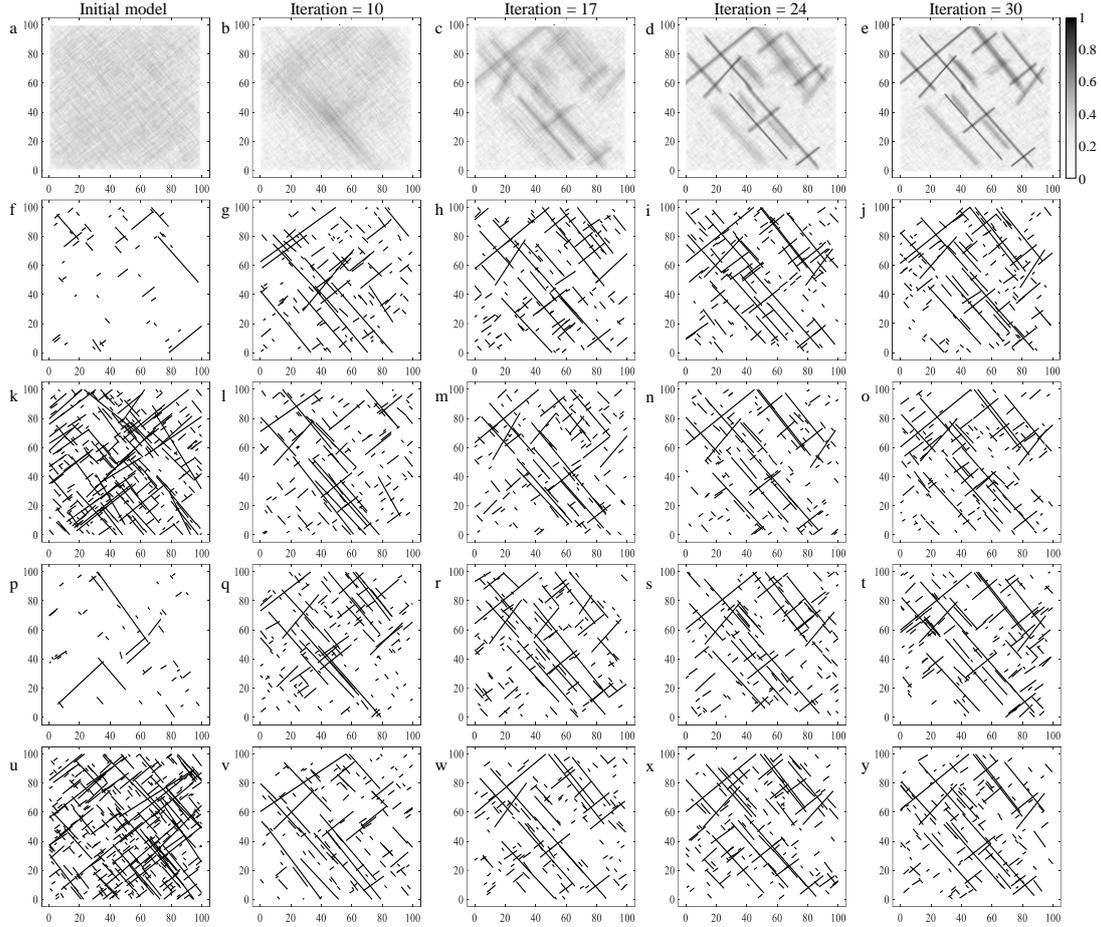

**Fig. 14.** The iterations of the fracture inversion process of the proposed algorithm for case 2. The rows are the fracture inversion process of (a-e) probabilistic map of fracture ensembles, and (f-y) randomly selected fracture distribution realizations, respectively. The columns are the initial model, after 10 iterations, after 17 iterations, after 24 iterations, and after 30 iterations during data assimilation process, respectively.

To validate the effectiveness of the proposed framework, ES is adopted in case 2 for comparison. **Fig. 13** indicates the iterations of the fracture inversion process of ES for case 2 using probabilistic map of fracture ensembles and randomly selected fracture realizations. **Fig. 14** presents the iterations of the fracture inversion process of the proposed algorithm for case 2. Before the inversion, the uncertainty of the prior realizations is relatively high. During the inversion process, the probabilistic map of fracture ensembles gradually presents the possible azimuth and the locations of large fractures. The fracture realizations of prior models of ES satisfy the constraints well.

After inversion, the approximate location of large fractures can be captured. Nevertheless, the fracture azimuth obtained by ES gradually deviates from two main orientations during the data assimilation process. In comparison with the true fracture distribution in **Fig. 11**, the inversion result of the proposed algorithm captured multiple locations of large fractures precisely. Besides, the azimuth of fractures obeys the prior constraints and mainly follows two orientations. A few fractures deviate the true location, which can be attributed to the multiple solutions of inverse problem. The probabilistic map of the updated results reveals significant reduction in the uncertainty of data matching.

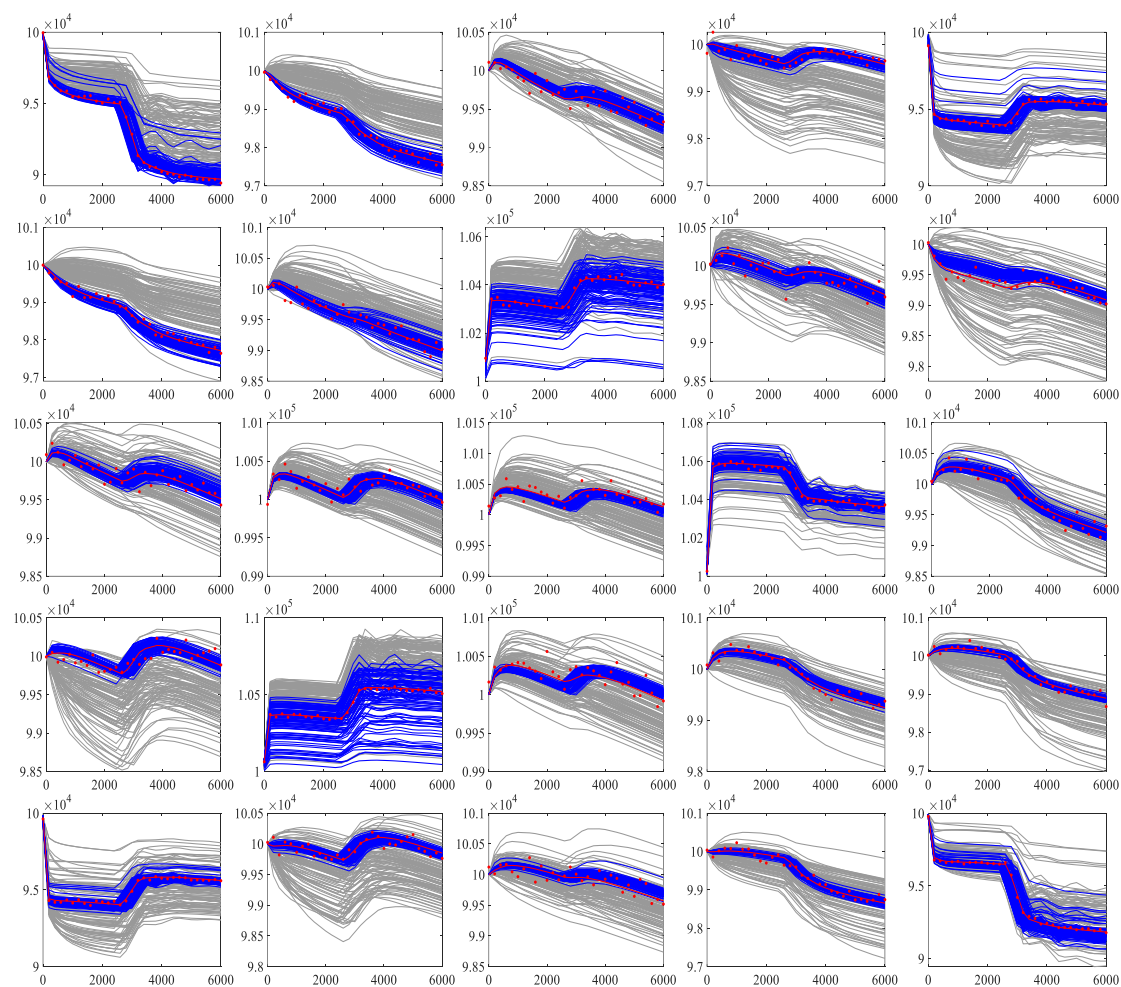

**Fig. 15.** The matching results of ES for case 2. The red curves are the prediction of true fracture field, and the red points are the observing data with noise. The gray and blue curves are the predictions of initial ensembles and updated ensembles.

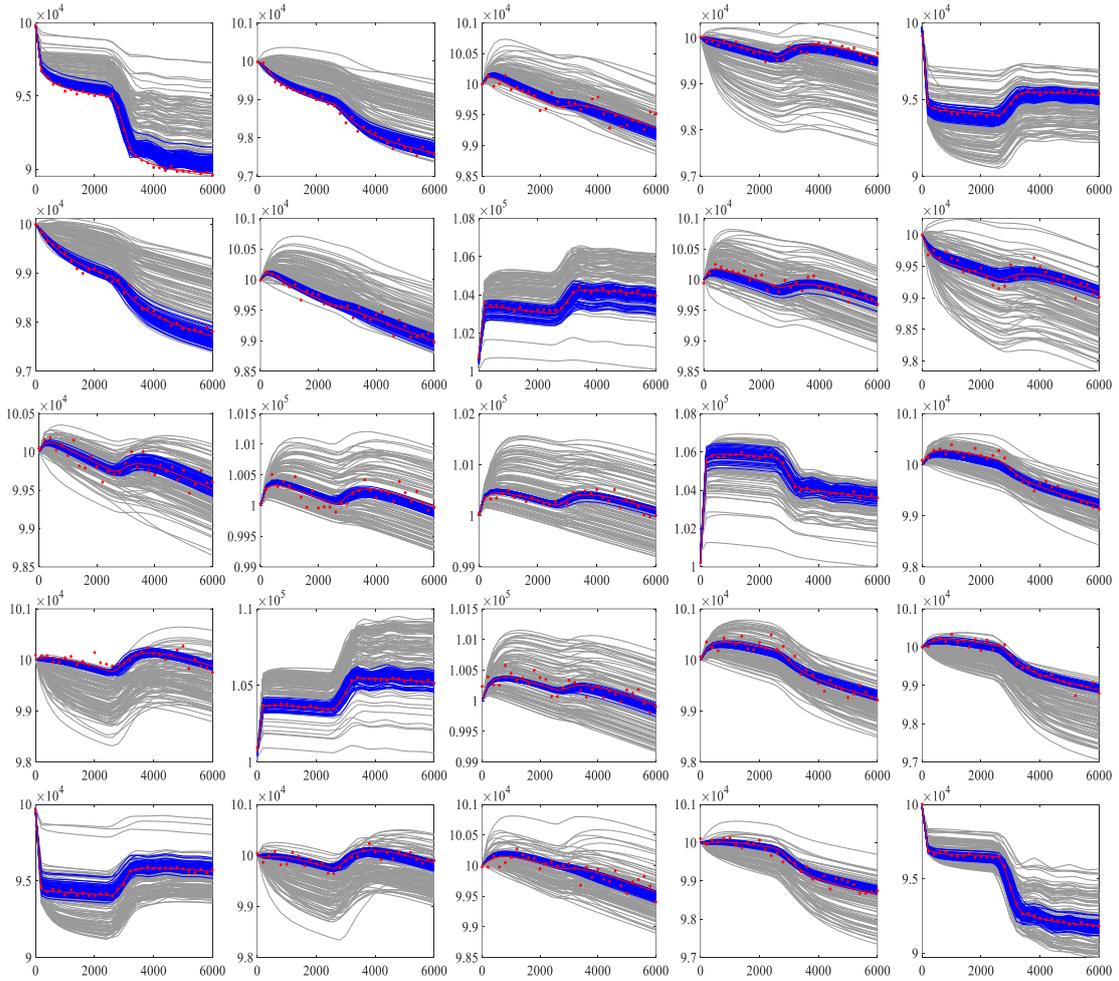

**Fig. 16.** The matching results of the proposed algorithm for case 2. The red curves are the prediction of true fracture field, and the red points are the observing data with noise. The gray and blue curves are the predictions of initial ensembles and updated ensembles.

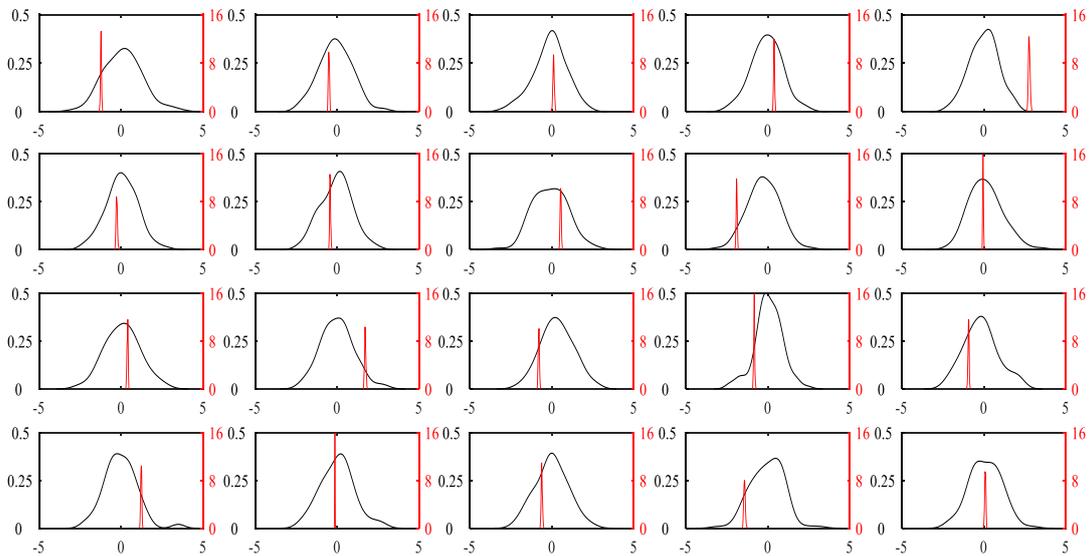

**Fig. 17.** Case 2: The prior distributions (black) of the latent variables for the initial ensemble and the posterior distributions (red) of the latent variables for the updated ensemble after data assimilation.

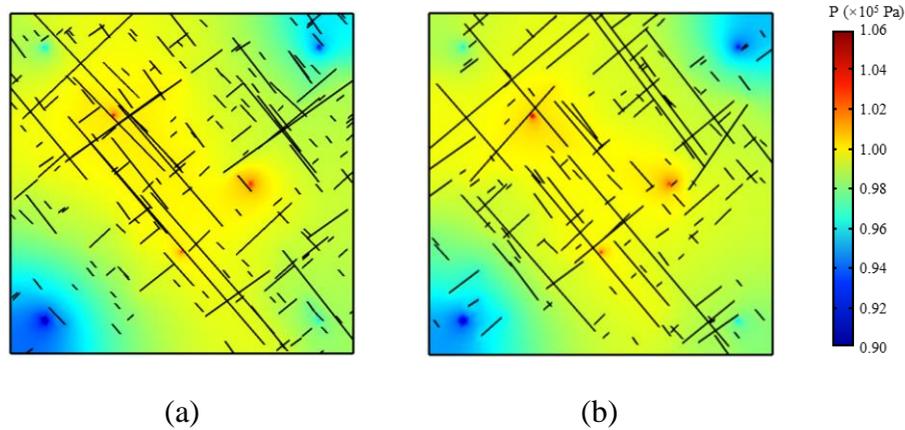

(a) (b)

**Fig. 18.** Case 2: The pressure prediction at 6000 s for (a) the reference model, and (b) a randomly selected posterior fracture realization.

**Figs.** 15 and 16 present the matching results of ES and the proposed algorithm in case 2, respectively. The matching result of ES is not as convergent as the proposed algorithm, resulting in that fracture realizations exhibit more variability. **Fig.** 17 illustrates the prior and posterior distributions of the 40 latent variables for the initial ensemble and updated ensemble after data assimilation, respectively. The prior distributions of the latent variables are standard normal distribution. After inversion, the posterior distributions of the latent variables converge to a normal distribution with small variation. The result of case 2 illustrates that there is not sufficient data for reliable fracture inversion. Although the observing data matched well, the updated fracture network ensemble is not close to the true fracture distribution. **Fig. 18** illustrates the pressure prediction for a randomly selected posterior fracture realization at 6000 s and the reference model. Although the positions of some large fractures do not find the accurate position, the estimated fracture model is able to match the main trend of the pressure distribution. **Fig. 19** presents that the fracture density and fractal dimension for case 2 can be properly identified.

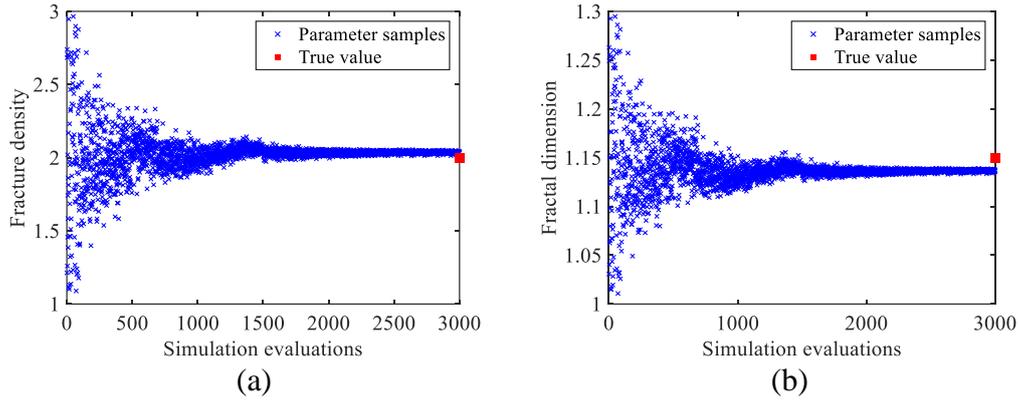

**Fig. 19.** Trace plots for the inversion parameters with the proposed algorithm (a) fracture density, and (b) fractal dimension.

## 5. Discussions

Estimating statistic parameters of fracture fields (i.e., fracture density and fractal dimension) is not trivial, since the same parameters may generate different fracture realizations. For small fractures, fracture density can merely quantify the total number of small fractures, and do not ideally describe the locations of each fracture. Besides, it is impossible to describe each small fracture in detail as the number of parameters would be huge. Therefore, the forward simulations of each vector of fracture parameters will have various predictions which increases the difficulty of fracture inverse modeling. In order to alleviate the impact of multiple realizations with the same parameters, average predictions of the observing data are employed by running multiple simulations.

Fracture distributions are diverse with different geological structures and distribution properties, which can be constrained by sufficient prior information. Although the applicability of the deep learning method to real-world geo-reservoir settings is limited by the type and amount of data requirements, deep learning based algorithm presents powerful performance on parameterization for the inference of complex geological structures and non-Gaussian fracture and permeability fields. This study considers two kinds of fracture distribution, i.e., one has designated number of large fractures in the prior information, and the second one has a complex fracture network containing both large and small fractures. How to use multifractal characterization for inverse modeling of fracture network to take into account the heterogeneity of fracture density is also a worthy research direction.

It is worth mentioning that the final updated ensemble tends to converge to similar fracture distribution pattern. To increase the diversity of the final solutions, multi-modal based techniques can be adopted, such as multimodal heuristic algorithms [65] and iterative local updating ensemble algorithms [22]. Correspondingly, more computational resources will be needed to explore more local or global optimal regions. Heuristic algorithms employ several populations, ensemble-based algorithms employ multiple sub-ensembles, and gradient based algorithms employ multiple start points to achieve diverse solutions for multimodal problems.

How to obtain optimal reservoir development scheme to achieve efficient heat extraction is of great significance for geothermal exploitation [66]. However, most existing optimization algorithms need to consume thousands of simulation evaluations [67]. Besides, field-scale geothermal simulation consumes hours even days. Thus, there is an urgent need to reduce simulation runs and improve the quality of the final development plan for geothermal reservoirs. The inverse modelling and the promote algorithm to reconstruct the fracture density and distribution in this study contributes to optimal geothermal development and decision-making by the shareholders to achieve efficient heat extraction.

## 6. Conclusion

For the nonlinearity and non-Gaussian distribution of fracture inversion problems, a novel deep learning based inverse modeling framework is proposed for estimating the fracture field parameters. For complex fracture network model, a hierarchical parameterization method is adopted. Each fracture is characterized specifically by length, azimuth and coordination of the fracture center for small number of large fractures, while fracture density and fractal dimension are utilized to characterize the fracture networks for dense small fractures. Subsequently, VAE and GAN are combined, and GAN objective is fused with prior constraint information to capture the distribution of the parameters of complex fracture networks and satisfy the prior constraint information, thereby mapping the high-dimensional complex parameter distribution into low-dimensional continuous parameter field. Moreover, ensemble smoother is used based on the observing data from hydraulic tomography to estimate the parameters of the fracture distribution.

To validate the effectiveness of the proposed deep learning based inversion framework on fractured flow problems, two numerical cases with different complexity are conducted. The results show the deep generative model is capable to capture the main features of the prior fracture network. The low-dimensional latent variables can map back to high-dimensional fracture parameter space and shows continuous change on high-dimensional space with continuous perturbation of low-dimensional latent variables. Besides, the proposed algorithm shows promising performance on estimating the distribution of the fracture fields after providing sufficient prior constraint information and hydraulic measurements.

## Acknowledgements

This study is supported by grants from the Research Grants Council of the Hong Kong Special Administrative Region, China (Project No. 17303519 and 17307620), and the HKU Seed Fund for Basic Research.